\documentclass[preprint,showpacs,preprintnumbers,amsmath,amssymb,prb]{revtex4}

\usepackage{graphicx}
\usepackage{dcolumn}
\usepackage{bm}

\begin{document}


\title{Inelastic scattering in a monolayer graphene sheet; a weak-localization study}

\author{Dong-Keun Ki}
 \thanks {These authors equally contributed to this work.}
 \affiliation{Department of Physics, Pohang University of Science
and Technology, Pohang 790-784, Korea}

\author{Dongchan Jeong}
 \thanks {These authors equally contributed to this work.}
 \affiliation{Department of Physics, Pohang University of Science
and Technology, Pohang 790-784, Korea}

\author{Jae-Hyun Choi}
 \affiliation{Department of Physics, Pohang University of Science
and Technology, Pohang 790-784, Korea}

\author{Kee-Su Park}
\affiliation{Department of Physics, Pusan National University,
Busan 609-735, Korea}

\author{Hu-Jong Lee}
\email{hjlee@postech.ac.kr} \affiliation{Department of Physics,
Pohang University of Science and Technology, Pohang 790-784,
Korea}

\date{\today}

\begin{abstract}
Charge carriers in a graphene sheet, a single layer of graphite,
exhibit much distinctive characteristics to those in other
two-dimensional electronic systems because of their chiral nature.
In this report, we focus on the observation of weak localization
in a graphene sheet exfoliated from a piece of natural graphite
and nano-patterned into a Hall-bar geometry. Much stronger
chiral-symmetry-breaking elastic intervalley scattering in our
graphene sheet restores the conventional weak localization. The
resulting carrier-density and temperature dependence of the phase
coherence length reveal that the electron-electron interaction
including a direct Coulomb interaction is the main inelastic
scattering factor while electron-hole puddles enhance the
inelastic scattering near the Dirac point.

\end{abstract}


\maketitle

\section{INTRODUCTION}

\begin{figure}[p]
\begin{center}
\leavevmode
\includegraphics[width=0.7\linewidth]{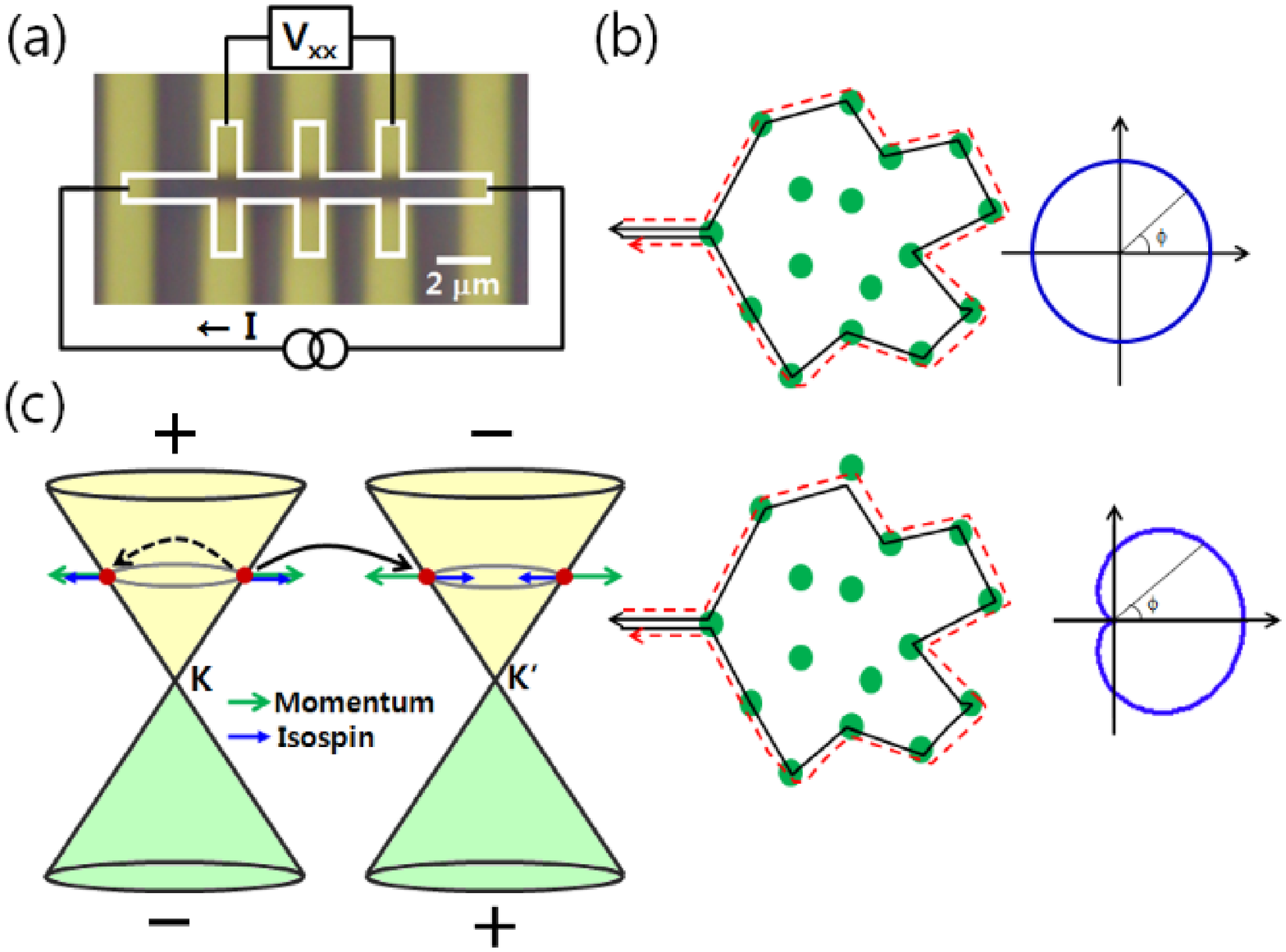}
\caption{\label{fig:sample} }
\end{center}
\end{figure}

Since charge carriers in graphene, a single layer of graphite,
have the unique chiral nature,~\cite{Geim07} the interference of
carriers~\cite{Russo08,Tikhonenko08,Wu07,Staley08,Graf07,Morozov06,
Heersche07} is not only sensitive to breaking of the phase
coherence represented by the characteristic scattering rate of
$\tau^{-1}_{\phi}$ but also sensitive to breaking of the chiral
symmetry.~\cite{Suzuura02,McCann06} In particular, in the case of
weak localization~\cite{Bergman84} (WL) in graphene, the random
gauge field generated at surface ripples~\cite{Meyer07} is
believed to suppress the WL effect~\cite{Morozov06} while
conventional WL~\cite{Tikhonenko08,Heersche07} and
anti-WL~\cite{Wu07} have also been reported. In general, WL of
charge carriers takes place as they are coherently
back-scattered~\cite{Bergman84} [Fig. \ref{fig:sample}(b)]. Charge
carriers in graphene, however, accumulate the Berry's phase $\pi$
if the chiral symmetry is conserved during the coherent back
scattering as along a gray-circle path on the surface of a
cone-shape dispersion relation in Fig.
\ref{fig:sample}(c).~\cite{Suzuura02,McCann06} Thus, as long as
the chiral symmetry is conserved, anti-WL is expected to be
revealed in graphene due to the sign change of the quantum
interference correction to the
conductivity.~\cite{Wu07,Suzuura02,McCann06} But, this anti-WL
itself is suppressed by the trigonal warping, which breaks the
time-reversal symmetry in a valley at the rate of $\tau^{-1}_w$,
and also by the chiral-symmetry-breaking elastic intravalley
scattering~\cite{Suzuura02,McCann06} as denoted by the dotted
arrow in Fig. \ref{fig:sample}(c) at the rate of $\tau^{-1}_z$.

On the other hand, the conventional character of WL can be
recovered by the elastic intervalley
scattering~\cite{Suzuura02,Tikhonenko08,McCann06} at the
characteristic rate of $\tau^{-1}_i$ [denoted by the solid arrow
in Fig. \ref{fig:sample}(c)]. Both the intravalley and the
intervalley scattering are possible only when the chiral symmetry
is broken by surface ripples, atomic-sized defects, boundaries,
etc.~\cite{Suzuura02} Thus, WL in graphene strongly depends on
such detailed sample conditions. This can be the reason for some
contradictory outcomes in previous
observations.~\cite{Morozov06,Wu07,Tikhonenko08,Heersche07}

We report WL in a Hall-bar-patterned 1-$\mu$m-wide graphene sheet
[denoted by a white line in Fig. \ref{fig:sample}(a)] with much
stronger elastic intervalley scattering ($\tau_i^{-1} \gg
\tau_{\phi}^{-1}$) than in previous
works.~\cite{Morozov06,Tikhonenko08,Wu07} Detailed examination of
the gate-voltage and temperature dependencies of the phase
coherence length ($L_{\phi}$) in this study indicates that the
electron-electron interaction~\cite{Altshuler85,Hansen93} is the
main inelastic scattering factor at temperatures of our
measurement range (below $\sim$20 K)~\cite{Tikhonenko08,Morozov06}
while electron-hole puddles~\cite{Martin08} enhance the inelastic
scattering near the Dirac point (DP).~\cite{Tikhonenko08,Staley08}

\section{SAMPLE PREPARATION AND MEASUREMENTS}

A mono-layer graphene sheet used in this study was mechanically
exfoliated~\cite{Geim07} from a natural graphite piece onto a
highly doped silicon substrate covered with a 300-nm-thick SiO$_2$
layer. After electric contacts were made by conventional e-beam
lithography and subsequent e-beam evaporation of Cr (5 nm) and Au
(70 nm), a Hall-bar geometry was defined by oxygen plasma etching
(for 90 seconds at 50 W) with a negative e-beam-resist stencil
patterned by e-beam lithography [denoted by a white line in Fig.
\ref{fig:sample}(a)].

\begin{figure}[p]
\begin{center}
\leavevmode
\includegraphics[width=0.7\linewidth]{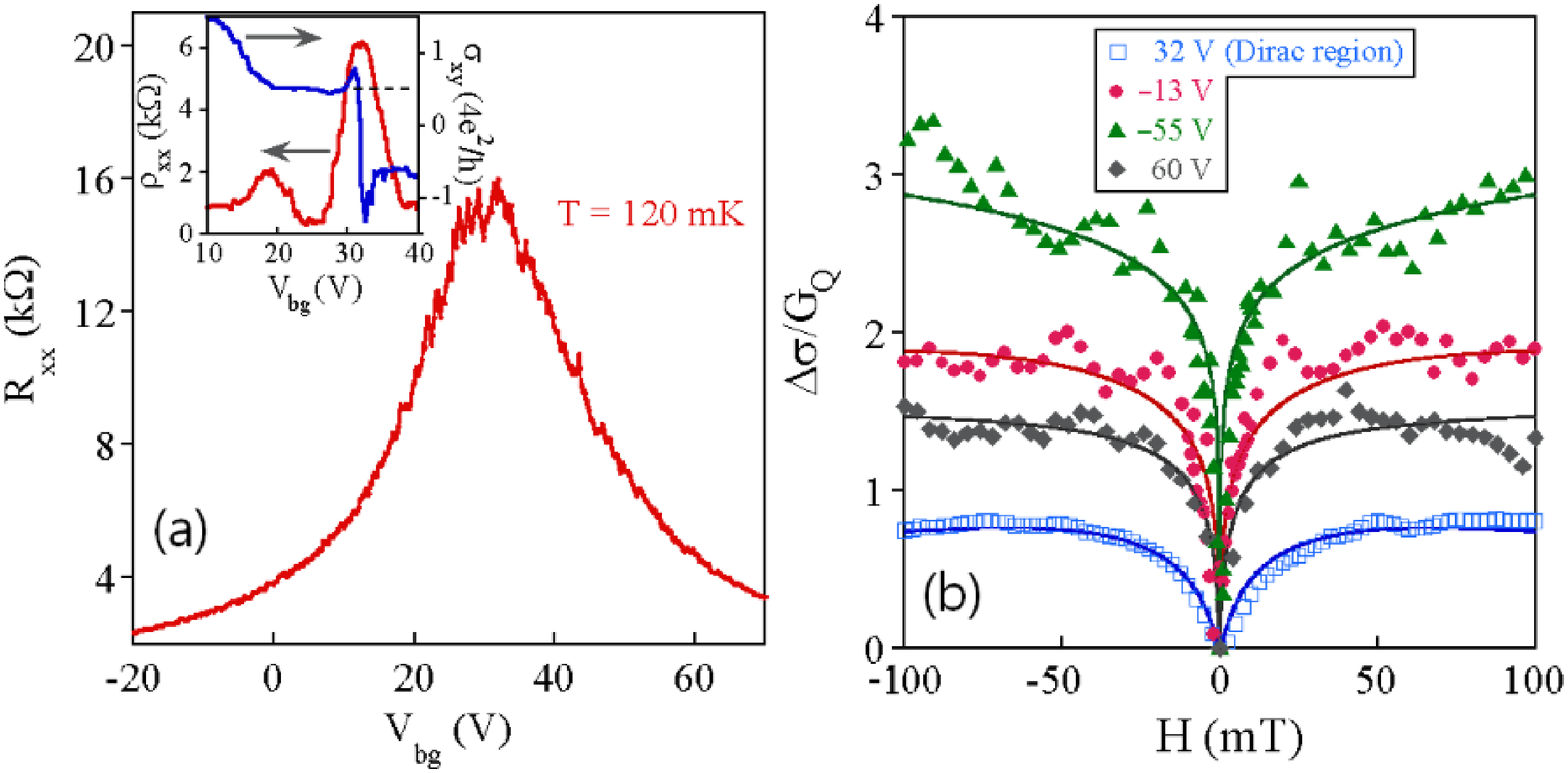}
\caption{\label{fig:datah} }
\end{center}
\end{figure}

Measurements were made by using a dilution fridge (Oxford Model
Kelvinox) with the base temperature as low as 120 mK. Each lead
contact showed resistance of $\sim$ 400 $\Omega$ with a negligible
temperature dependence. The longitudinal sample resistance vs. the
back-gate voltage as shown in Fig. \ref{fig:datah}(a) reveals the
charge-neutrality DP at $V_{bg}$=32 V. In addition, the
half-integer quantum Hall effect~\cite{Geim07} in a field of 9 T
[the inset of Fig. \ref{fig:datah}(a)] confirms the single
layeredness of our graphene sheet. Because of the mesoscopic size
of the sample ($1\times6$ $\mu$m$^2$), the universal conductance
fluctuation was also observed just like in the previous
studies~\cite{Staley08,Tikhonenko08,Morozov06,Heersche07,Graf07}
(not shown). Thus, following Ref. [\onlinecite{Tikhonenko08}], the
fluctuation in the magnetoconductivity (MC) was averaged out over
a 2 V range of back-gate voltage to get the conductivity
correction as $\Delta
\sigma(V_{bg},B)=[\sigma(V,B)-\sigma(V,0)]_{(V_{bg}-1 V \leq V
\leq V_{bg}+1 V)}$. The resultant data are shown as dots in Fig.
\ref{fig:datah}(b).

\section{RESULTS AND DISCUSSION}

For analysis, we use the expression for the WL-induced
conductivity correction as theoretically suggested with three
parameters~\cite{McCann06} ($L_{\phi}$, $L_i$, and $L_{\ast}$)

\begin{eqnarray}
\Delta \sigma =\frac{e^2}{\pi h} \times [\textrm{F}(\frac{8 \pi
B}{\Phi_0 L^{-2}_{\phi}})-\textrm{F}(\frac{8 \pi B}{\Phi_0
\{L^{-2}_{\phi}+2L^{-2}_{i}\}})-2\textrm{F}(\frac{8 \pi B}{\Phi_0
\{L^{-2}_{\phi}+L^{-2}_{i}+L^{-2}_{\ast}\}})],\label{eq:wl}
\end{eqnarray}
\\
\noindent where $\textrm{F}(z)=\textrm{ln}z+\psi(0.5+z^{-1})$,
[$\psi(x)$, the digamma function; $\Phi_0$(=$h/e$), flux quantum].
$L_{\phi}$ and $L_i$ stand for the phase coherence length and the
elastic intervalley scattering length with the relation
$L_{\phi,i}=\sqrt{D\tau_{\phi,i}}$ ($D$ is the diffusion constant
of the graphene sheet), respectively, while $L_{\ast}$ is defined
by the combination of $\tau^{-1}_z$ and $\tau^{-1}_w$ as,
$L_{\ast}=\sqrt{D\tau_{\ast}}$ and
$\tau^{-1}_{\ast}=\tau^{-1}_z+\tau^{-1}_w$.

Fig. \ref{fig:datah}(b) shows the differential MC data taken at
the base temperature of 120 mK for the selected gate voltages,
which show the positive MC, $\it{i.e.}$, the conventional
character of WL. Best fits of the data to Eq. (\ref{eq:wl}) are
shown as solid lines in Fig. \ref{fig:datah}(b). The conductivity
dip at zero magnetic field gets deeper as the gate voltage is
shifted from the DP, which indicates an increase of $L_{\phi}$
with increasing the carrier density $n$. Although similar behavior
was also observed previously~\cite{Tikhonenko08} at a few gate
voltages, no systematic studies on this effect were
pursued.~\cite{Graf07} The best-fit values for the parameters
$L_{\phi}$, $L_i$, and $L_{\ast}$, determined from the plot in
Fig. \ref{fig:datah}(b), are given in Fig. \ref{fig:gate}(a),
which clearly show that $L_{\phi}$ becomes longer as the carrier
density increases. However, the characteristic lengths for the
elastic intervalley scattering ($L_i$) and other kinds of elastic
scattering ($L_{\ast}$) remain almost unchanged at values much
smaller than $L_{\phi}$ at all gate voltages.

Since only the first term in Eq. (\ref{eq:wl}) gives a positive
MC, the conventional character of WL becomes most prominent when
both the intravalley and the intervalley scattering are strong so
that $L_{\phi} \gg L_i \gtrsim L_{\ast}$, which is the case of
this study. In comparison, the previous study by Morozov
\textit{et al.}~\cite{Morozov06} corresponds to the negligible
intervalley scattering as $L_i \gg L_{\phi} \gg L_{\ast}$, while
$L_{\phi} \sim L_i \gg L_{\ast}$ for the majority of the samples
in the work by Tikhonenko \textit{et al.}~\cite{Tikhonenko08} A
high intervalley scattering rate was also observed in the
narrowest sample of the work of Ref. [\onlinecite{Tikhonenko08}],
which was caused by the boundary scattering. By contrast, the
strong elastic intervalley scattering in our study takes place in
a much wider graphene sheet, where the boundary scattering is
supposedly negligible. Thus, we believe that the strong elastic
intervalley scattering in our system was mainly caused by
atomically sharp defects, which are known to make both the
intervalley and the intravalley scattering
strong.~\cite{Tikhonenko08} The sharp defects in our sample may
have resulted from the strong coupling between the graphene sheet
and the Si substrate or any mechanical defects introduced during
the exfoliation process.

As the conventional WL effect is restored by the high elastic
intervalley scattering rate in our graphene sheet, a convenient
condition is provided to examine the \emph{inelastic scattering}
characteristics in graphene. At low temperatures, the
electron-electron interaction~\cite{Altshuler85,Hansen93} is
considered to be the major source of the inelastic scattering in
graphene.~\cite{Tikhonenko08,Morozov06,Gonzalez96} The inelastic
scattering by the electron-electron interaction can be divided
into two terms; the direct Coulomb interaction ($\tau^{-1}_{ee}$)
among electrons and the interaction of an electron with the
fluctuating electromagnetic field generated by the noisy movement
of neighboring electrons (Nyquist scattering,
$\tau^{-1}_N$).~\cite{Hansen93,Altshuler85} Both interactions
depend on the carrier density, along with the screening by
neighboring charge carriers, which was proven a decade ago in a
GaAs/AlGaAs heterojunction two-dimensional electron gas (2DEG)
system.~\cite{Hansen93} In graphene, different from 2DEG, $E_F
\propto k_F \sim \sqrt{n}$ and the Fermi velocity $v_F$ is
independent of $n$,~\cite{Geim07} so that the interaction depends
on the carrier density as

\begin{eqnarray}
1/\tau_{\phi} &\varpropto& 1/\tau_N + 1/\tau_{ee} + \textrm{const}
\nonumber \\
&\varpropto& a k_BT \frac{\textrm{ln}(g)}{\hbar g}+b
\frac{\sqrt{\pi}}{2v_F}(\frac{k_BT}{\hbar})^2
\frac{\textrm{ln}(g)}{\sqrt{n}}+\textrm{const},\label{eq:eeg}
\end{eqnarray}
\\
\noindent where $g(n)$ is the normalized conductivity defined as
$g(n)=\sigma(n) h/e^2$ ($\sigma$; conductivity, $h$; Planck's
constant). The first term corresponds to the inelastic scattering
with a small momentum transfer. The direct Coulomb interaction,
represented by the second term, corresponds to a
large-momentum-transfer collision.~\cite{Hansen93,Altshuler85}
This term, negligible near the base temperature of this study,
shows a finite effect on the inelastic scattering at higher
temperatures below $\sim$20 K. This direct Coulomb interaction has
not been considered for the inelestic scattering in previous
works~\cite{Tikhonenko08,Morozov06,Wu07}.

To compare the data with the prediction of Eq. (\ref{eq:eeg}), the
inelastic scattering rate ($\tau^{-1}_{\phi}$) is estimated from
$L_{\phi}$ in Fig. \ref{fig:gate}(a) with the diffusion constant
$D$ obtained from the relations, $\sigma=(2e^2/h) k_F l$ and
$D=v_F l/2$. Here, since $n$ vanishes at the DP, we adopted the
value of $D$ at $V_{bg}$=31 V (just 1 V away from the DP) as its
value of the DP. The resultant values of $\tau^{-1}_{\phi}$ and
the best fit with the \emph{first term} in Eq. (\ref{eq:eeg}) are
shown as filled circles and a solid line, respectively, in Fig.
\ref{fig:gate}(b). The figure shows a nice fit for the gate
voltages beyond $\sim$15 V from the DP with $a \approx 10$ (for
clarity, the data at DP are not shown in the figure). The growing
discrepancy of the observed $\tau^{-1}_{\phi}$ from the fitting
approaching the DP strongly suggests that additional inelastic
scattering is present near the DP. As suggested in earlier
studies,~\cite{Tikhonenko08,Staley08} electron-hole
puddles~\cite{Martin08} can be the cause of this behavior.
Although the average carrier density vanishes near the DP,
electrons and holes may form spatially fluctuating regions called
puddles.~\cite{Martin08} In this circumstance, puddles can
generate fluctuating electromagnetic field which enhances the
Nyquist scattering (large $\tau^{-1}_N$). Thus, near the DP,
charge carriers are supposed to be inelastically scattered more
frequently than at gate voltages away from the DP, which can lead
to additional phase breaking near the DP.

Fig. \ref{fig:gate}(c) illustrates the temperature dependence of
the inelastic ($L_{\phi}$) and the elastic scattering lengths
($L_i$ and $L_{\ast}$) for the region of dense electron carriers
($V_{bg}$=60 V). It is seen that the relation $L_{\phi} \gg L_i
\gtrsim L_{\ast}$ holds in almost the whole low-temperature range
used in this study, where $L_i$ and $L_{\ast}$ turn out to be
almost temperature independent. The inset of Fig.
\ref{fig:gate}(d) shows the temperature dependence of $L_{\phi}$
at three gate voltages including the DP. In Figs.
\ref{fig:gate}(c) and \ref{fig:gate}(d), $L_{\phi}$ increases with
lowering temperature and is saturated at a certain value below a
characteristic temperature $T_{sat}$, as marked in the inset of
Fig. \ref{fig:gate}(d) by an arrow for each value of $V_{bg}$. The
saturation of $L_{\phi}$ is simply expected with a finite
concentration of magnetic impurities.~\cite{Pierre03} However, as
shown in the inset of Fig. \ref{fig:gate}(d), the value of
$T_{sat}$ depends on the carrier density (or $V_{bg}$). Since the
amount of magnetic impurities should be independent of the carrier
density, this behavior cannot be explained by the existence of
magnetic impurities. As an alternative explanation of this
saturation, one can resort to the formation of the electron-hole
puddles~\cite{Martin08} again. By the presence of the
electron-hole puddles, the effective conducting area for a charged
carrier is reduced,~\cite{Martin08} which acts to limit the
increase of $L_{\phi}$. Approaching closer the DP, with a higher
density of electron-hole puddles, the saturation of $L_{\phi}$
starts to take place at higher temperatures as shown in the inset
of Fig. \ref{fig:gate}(d).

In Fig. \ref{fig:gate}(d), the inelastic scattering rates
$\tau^{-1}_{\phi}$ at three gate voltages are shown as a function
of temperature, where values of $\tau^{-1}_{\phi}$ for $V_{bg}$=60
V (-35 V) is multiplied by a factor 3 (10) for clarity of the
temperature dependence. As denoted by the solid lines in Fig.
\ref{fig:gate}(d), data fit well to Eq. (\ref{eq:eeg}) at
temperatures higher than $T_{sat}$. Nice fits are obtained only
when the $T^2$ term in Eq. (\ref{eq:eeg}) is included, especially
as $V_{bg}$ deviates more from the DP. When we put Eq.
(\ref{eq:eeg}) as $\tau^{-1}_{\phi}=\alpha T+\beta T^2$ the ratio
$\beta/\alpha$ turns out to be about 0.016, 0.064, and 0.43 in an
arbitrary unit for $V_{bg}$ of 32 (the DP), 60, and -35 V,
respectively. The ratio, a reference of the relative strength of
the two terms in Eq. (\ref{eq:eeg}), indicates that the relative
importance of the direct Coulomb interaction represented by the
$T^2$ dependence of $\tau^{-1}_{\phi}$ increases for a higher
$V_{bg}$, although the absolute scattering rates of both of the
two terms decrease.

In previous studies,~\cite{Tikhonenko08,Morozov06} data were
fitted with the $T$-linear term only. Thus, the fit in Fig.
\ref{fig:gate}(d) indicates that the direct Coulomb interaction is
stronger in our graphene sheet. The direct Coulomb interaction
involves a large momentum transfer,~\cite{Hansen93,Altshuler85} as
in the elastic intervalley scattering denoted by the solid arrow
in Fig. \ref{fig:sample}(c).~\cite{Suzuura02} This leads us to
assuming that atomically sharp defects also cause the finite
direct Coulomb interaction in our graphene sheet. Thus, in
addition to the elastic intervalley scattering, it is probable
that two charge carriers occupying states within the thermal
energy difference of $k_BT$ in different valleys are scattered to
each other in the vicinity of atomically sharp defects [the inset
of Fig. \ref{fig:gate}(b)]. Interestingly, assuming that the
scattering probability of a carrier in a valley is proportional to
the number of available states to be scattered into in another
valley, the cross-section of this \emph{inelastic
electron-electron intervalley scattering}, a two-particle process,
would have $T^2$ dependence. This is in accordance with the
$T^2$-dependence term in Eq. (\ref{eq:eeg}). The enhanced
inelastic intervalley scattering in our sample, presumably with
higher concentration of atomic-scale defects, also explains the
difference of the WL behavior between this study and the previous
ones.~\cite{Tikhonenko08,Morozov06} The existence of this
inelastic scattering mechanism may be further confirmed by
separate experimental means, such as photoemission spectroscopy.

The saturation of the WL correction of the conductance may be
caused as the phase coherence length $L_{\phi}$ becomes comparable
to the sample length with lowering
temperature.~\cite{Tikhonenko08} In our measurements, however, the
saturation value of $L_{\phi}$ sensitively varies with $V_{bg}$
and the saturation takes place for most range of $V_{bg}$ in this
study at temperatures far before $L_{\phi}$ reaches the sample
length. At $V_{bg}$=60 V, for instance, where electron-hole
puddles are barely present, $L_{\phi}$ is saturated at $\sim$ 1.8
$\mu$m [see the inset of Fig. \ref{fig:gate}(d)]. This indicates
that the finite size effect is not the main cause of the
saturation of $L_{\phi}$.

\section{SUMMARY}

In summary, we have measured the WL effect in graphene at various
gate voltages (or alternatively carrier densities) and
temperatures. The large chirality-breaking elastic intervalley
scattering, presumably from atomically sharp
defects,~\cite{Suzuura02} leads to relative characteristic length
scales of $L_{\phi} \gg L_i \gtrsim L_{\ast}$, which is in clear
distinction from the conditions of previous
works.~\cite{Morozov06,Wu07,Tikhonenko08,Heersche07} It points
that the transport properties in our graphene sheet may be more
plagued by extrinsic characters. However, the chirality-breaking
elastic intervalley scattering restores the conventional WL
effect.~\cite{Tikhonenko08,McCann06} The observed conventional WL
characters in our graphene are solely dependent on the
phase-breaking inelastic scattering~\cite{Bergman84} and thus
provide a convenient condition to examine the inelastic scattering
properties in graphene.

\begin{figure}[p]
\begin{center}
\leavevmode
\includegraphics[width=0.7\linewidth]{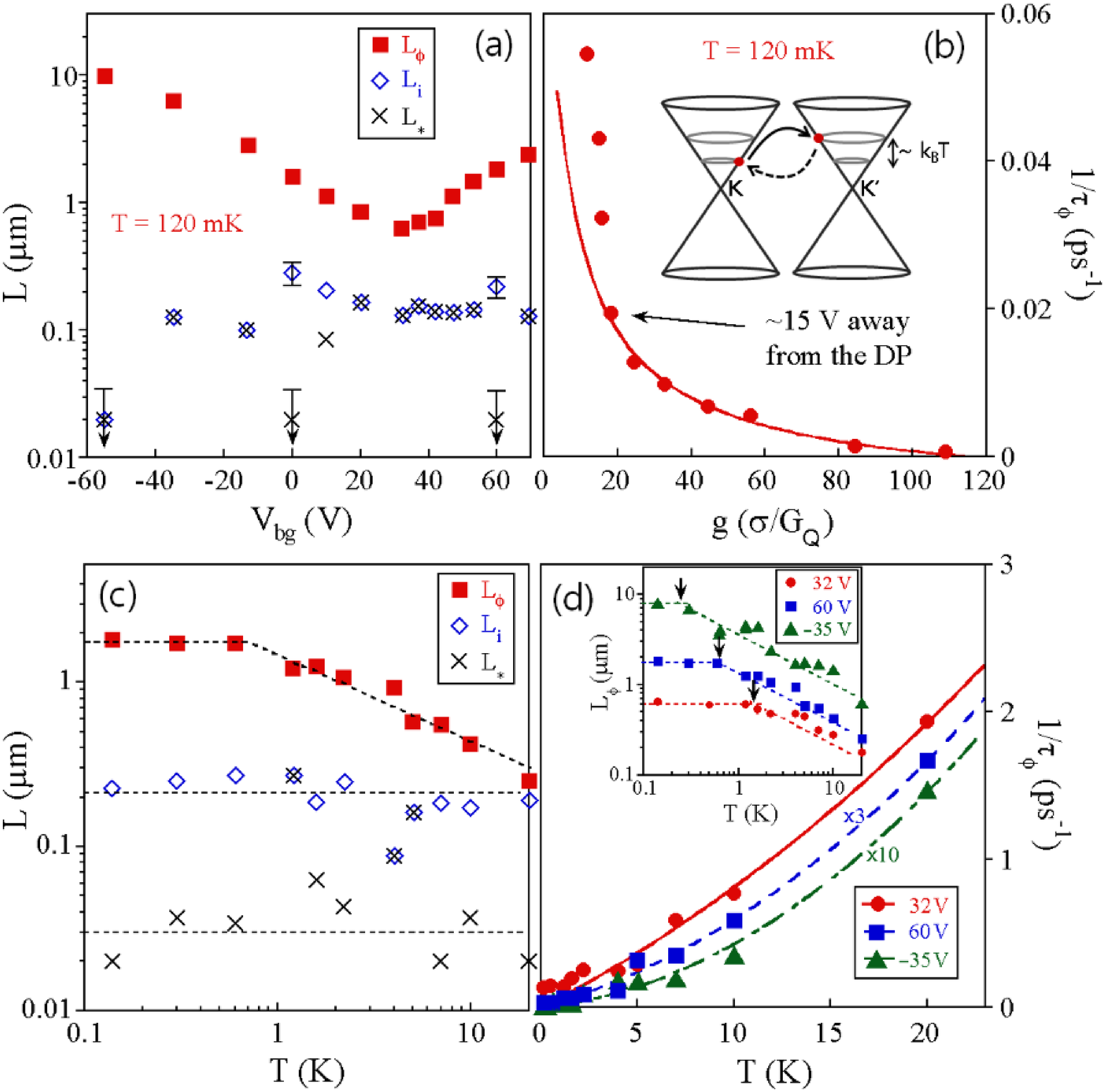}
\caption{\label{fig:gate} }
\end{center}
\end{figure}

In consequence, we could study the gate-voltage and temperature
dependencies of the phase coherence length in detail as shown in
Fig. \ref{fig:gate}. Similar dependencies have also been observed
in measurements of WL,~\cite{Tikhonenko08,Graf07}
UCF,~\cite{Staley08,Graf07} and Aharonov-Bohm
oscillation~\cite{Russo08}, but without the quantitative analysis.
In this study, we show that the unusual gate-voltage and
temperature dependencies of the phase coherence length are
quantitatively explained in terms of the dominant inelastic
electron-electron scattering,~\cite{Altshuler85,Hansen93} while
electron-hole puddles~\cite{Martin08} are expected to produce an
additional inelastic scattering near the
DP.~\cite{Tikhonenko08,Staley08} Different from previous
works,~\cite{Morozov06,Wu07,Tikhonenko08} the direct Coulomb
interaction becomes effective at high temperatures of our study.
We propose that it is caused by the large-momentum-transfer
intervalley inelastic scattering in the vicinity of atomically
sharp defects of charge carriers that occupy states within the
thermal energy difference of $k_BT$ in different valleys.

\begin{acknowledgments}
This work was supported by Acceleration Research Grant
R17-2008-007-01001-0 and Pure Basic Research Grant
R01-2006-000-11248-0 administered by Korea Science and Engineering
Foundation, and by Korea Research Foundation Grant
KRF-2005-070-C00055.
\end{acknowledgments}

\newpage
\textbf{FIGURE CAPTIONS}
\\
\\
Figure 1. (color online) (a) Sample geometry and measurement
configuration. The boundary of our graphene sheet is denoted by a
white line. (b) Upper set: time-reversal closed paths contributing
to WL in a normal electronic system with an isotropic scattering
probability (shown on the right). Lower set: similar time-reversal
closed paths in graphene with an anisotrpic scattering probability
(shown on the right). (c) A cone-shape energy dispersion relation
of graphene. There are two different K points (valleys) with
different chiralities (denoted by $\pm$). The dotted and solid
arrows represent the intravalley and the intervalley scattering,
respectively.
\\
\\
Figure 2. (color online) (a) The back-gate-voltage dependence of
the longitudinal sample resistance. The mobility ($\mu$) of the
sample is $\sim$ 4000 cm$^2$/Vs outside the Dirac region. Inset:
the half-integer quantum Hall effect measured in 9 T and at 120 mK
confirms that our graphene sheet is single-layered. (b) Correction
of the magnetoconductivity for various gate voltages ($G_Q =
e^2/h$). The carrier density can be estimated with a relation
$n$[in cm$^{-2}$]=$7\times10^{10}(V_{bg}$[in V]-32) with a
positive (negative) value for electrons (holes). Dots and lines
represent experimental data and best fits, respectively.
\\
\\
Figure 3. (color online) (a) Best-fit values of the characteristic
lengths as a function of gate voltage at the base temperature of
120 mK. As denoted by arrows, the lower bounds of the error bars
of $L_\ast$ for a few $V_{bg}$ go as low as $\sim$1 nm, which are
not physical and are considered to arise from the uncertainty
involved in fitting to the particular functional form of Eq.
(\ref{eq:wl}) as discussed in Ref. [3]. (b) Inelastic scattering
rate as a function of conductance $g$ at the base temperature, 120
mK. The line is a best fit to the first term in Eq.
(\ref{eq:eeg}). (c) Temperature dependence of the characteristic
lengths at a dense-electron region ($V_{bg}$$\sim$60 V). Dotted
lines are guides to the eyes. (d) Inelastic scattering rates at
different gate voltages as a function of temperature. The values
of $\tau^{-1}_{\phi}$ for $V_{bg}$=60 V (-35 V) is multiplied by a
factor 3 (10) for clarity of the temperature dependence. Lines are
best fits to Eq. (\ref{eq:eeg}). Inset: Temperature dependence of
$L_{\phi}$. Dotted lines are guide to the eyes and the arrows
indicate the saturation temperature ($T_{sat}$).

\newpage

\newpage

\newpage

\end{document}